\documentstyle[twocolumn,aps]{revtex}

\input{epsf}
\begin{document}

\draft
\title{Quantum fluctuations of coupled dark solitons
in trapped Bose-Einstein condensates}
\author{C. K. Law, P. T. Leung and M.-C. Chu}
\address{{Department of Physics,
The Chinese University of Hong Kong,}\\
{Shatin, NT, Hong Kong, China}}
\date{\today}
\maketitle

\begin{abstract}

We show that the quantum fluctuations associated with the
Bogoliubov quasiparticle vacuum can be strongly concentrated
inside dark solitons in a trapped Bose Einstein condensate. We
identify a finite number of anomalous modes that are responsible
for such quantum phenomena. The fluctuations in these anomalous
modes correspond to the `zero-point' oscillations in coupled dark
solitons.

\end{abstract}
\vspace{10mm} 

Dark solitons in quasi one-dimensional Bose-Einstein condensates
(BEC) are coherent structures created from macroscopic
excitations. Recent experiments have realized such dark solitons
by using the phase imprinting technique \cite{expt1,expt2}, and
several theoretical studies have analyzed the stability and
dynamical properties \cite{muryshev,fedichev,busch,feder,carr}, as
well as novel soliton-pair solutions in two-component Bose
condensates \cite{santos,anglin}. Like other familiar topological
states in BEC, dark solitons are described by macroscopic
wavefunctions. Even though atoms are excited, most of them still
share the same wavefunction. Such a quantum degeneracy is the
basis of treating the many-particle system as a single coherent
matter wave. This raises a fundamental issue about the robustness
of quantum degeneracy as a BEC is excited to higher energy states
that typically contain more than one solitons. The presence of
multiple dark solitons and their mutual interaction would
introduce extra quantum fluctuations affecting the macroscopic
coherence. In this paper we investigate the mechanism of such
quantum decoherence effects.

We approach the problem by studying the quantum fluctuations
associated with a chain of dark solitons in a trapped condensate.
Specifically, we are interested in the minimal fluctuations
represented by the {\em quasiparticle vacuum} in the Bogoliubov
theory. In the case of ground state condensates, the corresponding
quasiparticle vacuum leads to a cloud of incoherent atoms (usually
called condensate depletion) not described by the Gross-Pitaevskii
mean field equation \cite{singh}. Such incoherent atoms are
typically a small fraction of the condensate, their
effects can be ignored in most situations involving dilute
condensates. However, we find that the quasiparticle vacuum
associated with the dark solitons has distinctive features
accessible for experimental observations. In this paper we
show that incoherent atoms can be strongly concentrated at the
notches, and in addition we discover that this phenomenon is
mainly the quantum fluctuations in a finite number of collective
modes. The particle-like behavior of dark solitons is the key to
interpret our findings. We shall see how dark solitons under
confinement couple together as particles to execute collective
oscillations. Quantum uncertainties in both the positions and
momenta impose a fundamental limit of the degree of `darkness'
inside the dark solitons.

To begin we consider a Bose-Einstein condensate of weakly
interacting atoms confined in a harmonic potential. A quasi
one-dimensional condensate can be achieved in a trap in which the
transverse motion is tightly confined (frozen) and only the
longitudinal motion remains active. Given that $m$ is the atomic
mass and $\omega_T$ is the trap frequency, the second quantized
system is modeled by the Hamiltonian (in units of $\hbar
\omega_T$),
\begin{equation}
H={1 \over 2}\int_{}^{} {dx\hat \Psi ^{\dag}\left ( {-{{\partial
^2} \over {\partial x^2}}+x^2+ \bar \eta \hat \Psi ^{\dag} \hat
\Psi } \right)}\hat \Psi.
\end{equation}
Here the spatial coordinate $x$ is in units of $\left( {\hbar
/m\omega_T } \right)^{1/2} $, and $\bar \eta$ is the interaction
strength between atoms averaged over the transverse area
\cite{eta}. As usual, the field operator $\hat \Psi (x,t)$ can be
decomposed into a dominant coherent part and a fluctuation part,
i.e., $\hat \Psi = \sqrt N \Phi + \hat \phi $, where $N$ is the
number of condensate atoms. The mean field wavefunction $\Phi =
\Phi_n e^{-i \mu_n t}$ is the $n$-th excited state of the BEC
satisfying the nonlinear Schr\"odinger eigen-equation,
\begin{equation}
 - {1 \over 2}{{\partial ^2 } \over {\partial x^2 }}\Phi_n
 +{1 \over 2} x^2
 \Phi_n+ g \left| {\Phi_n} \right|^2 \Phi_n  = \mu_n
\Phi_n
\end{equation}
where $\mu_n$ is the eigen-energy and $g\equiv N \bar \eta$. The
quantum number $n$ is the number of nodes appearing in $\Phi_n$.
Therefore $\Phi_n$ represents a static $n$-dark soliton solution.

In Fig. 1, we show the typical shapes of $\Phi_n$ obtained from
exact numerical solutions for the case $g=200$. Although there are
no simple analytic expressions for $\Phi_n$, we find that
variational methods are useful to approximate these wavefunctions.
We construct the trial wavefunction as a product of single dark
soliton solutions embedded in a smooth background with the ground
state profile, i.e.,
\begin{equation}
\Phi _n (x) \approx \alpha_n \Phi _0 (x)\prod\limits_{j = 1}^n
{\tanh \left( {{{x - q_j} \over l_0} } \right)}.
\end{equation}
Here $\alpha_n$ is a normalization constant, $q_j$ is the position
of the $j$-th dark soliton, and $l_0^{-1} \equiv g^{1/2} \Phi _0
(0)$ is the inverse of the healing length (neglecting the slight
position dependence near the trap center). The ground state
wavefunction $\Phi _0 (x)$ can be further approximated by $g^{ -
1/2} \left( {\mu _0 - x^2 /2} \right)^{1/2} $ in the Thomas-Fermi
limit $g \gg 1$. We treat $q_j$ as variation parameters such that
the mean field energy
\begin{equation}
E_n = {N \over 2}\int {dx\left( {\left| {{{\partial \Phi _n }
\over {\partial x}}} \right|^2  + x^2 \left| {\Phi _n } \right|^2
+ g\left| {\Phi _n } \right|^4 } \right)}
\end{equation}
is stationary. With the trial wavefunction given in Eq. (3), the
$q_j$ obtained have about $1 \%$ error compared with the exact
numerical results in Fig. 2. Such a good agreement is understood
from the fact that dark solitons are sufficiently separated from
each other by several healing lengths $l_0$, which allows the
product form in (3) to be a close approximation in the
Thomas-Fermi limit. Indeed, we find that the separation between
the solitons is between 3$l_0$ to 6$l_0$ for a wide range of
interaction strengths $10^2 < g < 10^3$. We remark that dark
solitons repel each other \cite{repel}. The mechanical equilibrium
is attained by the inhomogeneous condensate background that
provides a confinement `force' holding the solitons together
\cite{busch}.

Having described the $n$-dark soliton mean-field solution, we now
proceed to discuss the fluctuation part $\hat \phi_n = \hat \Psi
-\sqrt N \Phi_n e^{-i \mu_n t}$. Following the standard Bogoliubov method
\cite{fetter}, we write $\hat \phi_n$ as,
\begin{equation}
\hat \phi_n (x,t) = \sum\limits_j^{} {c_j ^{(n)}} u_j^{(n)} (x)e^{
- i\omega _j t} + c_j^{(n)\dag}  v_j^{(n)*} (x)e^{i\omega _j t}.
\end{equation}
Here $c_j^{(n)\dag}$ and $c_j^{(n)}$ are quasi-particle creation
and annihilation operators associated with the $j$-th collective
mode. The mode functions $u_j^{(n)}(x)$, $v_j^{(n)}(x)$ and
frequencies $\omega_j$ are determined by
\begin{eqnarray}
{\cal L} u_j^{(n)} + g\Phi _n^2 v_j^{(n)}  &=& \omega _j u_j^{(n)}
\\
{\cal L} v_j^{(n)} + g\Phi _n^{*2} u_j^{(n)} &=& - \omega _j
v_j^{(n)}
\end{eqnarray}
where ${\cal L} \equiv {1 \over 2}\left( { - {{d^2 } \over {dx^2
}} + x^2 + 4g\left| {\Phi _n } \right|^2  - 2\mu _n } \right)$,
under the restriction: $\int {dx( {| {u_j^{(n)} }|^2 - |
{v_j^{(n)} }|^2 } )}  = 1 $. Using the expansion given in (5), the
Hamiltonian (1) is approximated by a diagonalized form: $H \approx
E_n + \sum_j {\omega _j^{} } c_j^{(n)\dag} c_j^{(n)}$, where cubic
and higher terms of $\hat \phi_n$ are ignored.

Our task is to determine the incoherent part of atom density
\begin{equation}
\rho_n (x,t) \equiv \left\langle {\hat \phi_n ^{\dag}(x,t) \hat
\phi_n (x,t)} \right\rangle
\end{equation}
associated with the vacuum state in which there are no
quasiparticles in all collective modes, i.e., $ \left\langle {}
\right. c_j^{(n)\dag} c_j^{(n)} \left. {} \right\rangle = 0 $. It
is not difficult to show that among all physical states, the
vacuum gives the minimal time-averaged $\rho_n(x,t)$.
With the help of Eq. (5), $\rho_n(x,t)$
arising from the vacuum fluctuations is given by: $\rho_n (x)
=\sum_j {| {v_j^{(n)}(x)} |^2}$. An integration of $\rho_n (x)$
over $x$ gives the total number of incoherent atoms $\delta N$,
i.e., $\delta N = \sum_j {\delta N_j } $, with $ \delta N_j = \int
dx | {v_j^{(n)} (x)}| ^2 $.

We solve Eqs. (6) and (7) numerically for the first thousand
collective modes in order to study the distribution of incoherent
atoms. Although $\delta N$ involves all collective modes, we find
that only the low-frequency modes are significant. This is shown
in Fig. 2 in which $\delta N_j$ as a function of $\omega_j$ is
plotted. For the cases with soliton numbers up to four, we see
that $\delta N_j$ diminishes quickly as the frequency increases.
If we superimpose the four cases in Fig. 2 together, the four
tails associated with the high frequency part of $\delta N_j$ all
roughly lie on the same curve. This implies that the presence of
more dark solitons essentially do not cause extra fluctuations in
the high frequency modes.

From now on we will focus on low frequency modes. One prominent
feature we discover is that {\em the number of modes with negative
frequencies equals the number of dark solitons}. These particular
modes are shown in Fig. 2 as black filled circles. In the
Bogoliubov theory, collective modes with negative frequencies and
positive norm (i.e., $\int {dx( {| {u_j^{(n)} }|^2 - | {v_j^{(n)}
}|^2 } )} = 1 $) are known as `anomalous modes'. Previous studies
on a single dark soliton \cite{feder} and a single vortex
\cite{anomalous} have identified the particle-like motion
associated with anomalous modes. For example, in the $n=1$ dark
soliton case, the anomalous mode describes the small amplitude
oscillations of the soliton at a frequency $-\omega_T/ \sqrt 2$ in
the Thomas-Fermi limit \cite{muryshev,busch,feder}. This agrees
with the exact anomalous mode frequency shown in Fig. 2a within
1.3 percent. By increasing the interaction strength $g$, we find
that the numerical frequency approaches $-\omega_T/ \sqrt 2$.

In Fig. 3 the incoherent atom density $\rho_n (x)$ is plotted. We
see that $\rho_n (x)$ (solid lines) are strongly concentrated at
the positions of the dark solitons. In each figure the
contribution from the anomalous modes is shown by the dash line,
which is the partial sum in Eq. (9), counting only the finite
number of modes with $\omega_j <0$. By comparing the curves, we
conclude that  {\em incoherent atoms near the dark solitons mainly
originate from the vacuum fluctuations in the anomalous modes}. In
fact, if we remove the background effect by defining the quantity
$\Delta (x) \equiv \rho_n(x) - \rho_0(x)$ as a measure of
fluctuations due to the presence of dark solitons, we find a more
striking evidence. This is shown in the inset of Fig. 3 in which
the solid lines ($\Delta (x)$) and the dash lines (from the
anomalous modes) are almost indistinguishable from each other.

Having elucidated the significance of anomalous modes, we now make
a connection to the particle-like motion of the dark solitons. The
fact that the number of anomalous modes and the number of dark
solitons are equal (see Fig. 2) provides an important clue. We can
imagine the $n$ mutually repulsive dark solitons, each being
pushed toward the trap center by the inhomogeneous background,
form a one-dimensional `lattice' at equilibrium. As a result,
there are $n$ normal modes of oscillations when the equilibrium
system is slightly disturbed. These modes correspond to the
anomalous modes that are well localized around the dark solitons.
For example, the frequency $\approx -\omega_T/ \sqrt 2$ appears in
all four cases in Fig. 2. We find that this mode corresponds to
the in-phase mode that the $n$ solitons move together without
changing their relative separations.

To elaborate the physical picture quantitatively, we present a
method to estimate the effective masses and frequencies of the
coupled oscillations of solitons. For definiteness, let us examine
the $n=2$ case, and consider the ansatz wavefunction of the dark
soliton pair,
\begin{equation}
\tilde \Phi_2 (x,t)= \tilde \alpha \Phi _0 (x) \psi_1 (x,t) \psi_2
(x,t).
\end{equation}
Here $\tilde \alpha$ is a normalization constant and $\psi
_j(x,t)$ is the exact solution of a dark soliton at the position
$q_j$ moving at a speed $\dot q_j$ $(j=1,2)$ in  a homogeneous
background, i.e.,
\begin{equation}
\psi _j(x,t)=i{{\dot q_j} \over v_s}+\sqrt {1-{{\dot q_j^2} \over
{v_s^2}}}\tanh \left[ {\sqrt {1-{{\dot q_j^2} \over
{v_s^2}}}\left( {{x-q_j}\over l_0} \right)} \right],
\end{equation}
where $v_s =g^{1/2} \Phi _0 (0)$ is the local speed of sound for
the ground state at the origin . The construction of $\tilde
\Phi_2 (x,t)$ generalizes the trial wavefunction (3) by including
the speed $\dot q_j$ as variational parameters. Our strategy is to
determine how the energy of the system changes if $q_j$ and $\dot
q_j$ slightly deviate from the stationary point: $\dot q_j=0$ and
$q_1=-q_2=Q$. This can be done by inserting the ansatz (9) into
Eq. (4) and keep only the quadratic terms. The energy change can
be cast in the form of two harmonic oscillators,
\begin{equation}
\Delta E\approx {1 \over 2}M \left( {\dot q_+^2 + \dot
q_-^2+\omega _+^2q_ +^2+\omega _-^2q_-^2} \right).
\end{equation}
Here we have defined the generalized coordinates for the in-phase
mode $(+)$ and the out-of-phase mode $(-)$ by: $q_{\pm}=(\delta
q_1 \pm \delta q_2)/ \sqrt 2$ with $\delta q_1= q_1 - Q$ and
$\delta q_2= q_2 + Q$.

We find that  $\omega_{\pm}$ obtained in Eq. (11) match quite well
with the exact numerical values of the anomalous mode frequencies,
noticing that ansatz (9) is not an exact solution. For example,
$\omega_{+}$ and $\omega_{-}$ attain $93 \%$ and $96 \%$ accuracy
respectively for the case with $g=200$. The effective mass $M$ for
each soliton turns out to be negative, which is a basic property
of dark solitons discussed in literature \cite{fedichev}. When $g
\gg 1$ is sufficiently large, we find $M \approx - 4 N \Psi_0(0)^2
l_0 \approx -3.2 N g^{-2/3}$ (in units of the atom mass $m$).
Since $g=N \bar \eta$ is proportional to $N$, the absolute value
of the soliton mass $M$ increases as $N^{1/3}$ as the number of
particles increases.

Treating the positions $q_j$ and momenta $p_j \equiv M \dot q_j$
as dynamical variables, we construct a {\em two-particle}
Hamiltonian based on the energy (11). The quantized form of the
Hamiltonian obtained corresponds to the part of the Bogoliubov
Hamiltonian containing the two anomalous modes. Therefore the
creation and annihilation operators associated with the anomalous
modes can be explicitly related to the position and momentum
operators $\hat q_j$ and $\hat p_j$. In this way the vacuum state
of quasiparticles in the anomalous modes are now recognized as the
zero-point oscillations of the coupled dark solitons \cite{zero}.

For the case $n=2$ discussed here, the vacuum state gives the
position fluctuations $ \left\langle {\Delta q_ \pm ^2 }
\right\rangle = 1/2M \omega _ \pm $. Since the width of each dark
solitons is about $l_0$, the fluctuations become significant near the
notches if the ratio $ \left\langle {\Delta q_j } \right\rangle
/l_0 $ ($j=1,2$) is greater or comparable to one. We find that
this occurs if $N^{1/6} \bar \eta ^{2/3}  > 1 $, according to the
expression of $M$ obtained above and the fact that $\omega_{\pm}$
are order one. We can estimate the number density of incoherent
atoms at the notches. For example, by counting only the
contributions from the two anomalous modes, we have \cite{remark}
\begin{equation}
\hat \phi_2 (x = \pm Q ) \approx \sqrt {{N}} \tilde \alpha {{\Phi
_0 (0 )}\over \sqrt 2} \left( {i{{\hat p_ +   \pm \hat p_ - }
\over {M v_s}} - {{\hat q_ + \pm \hat q_ - } \over {l_0 }}}
\right).
\end{equation}
Therefore the density $\left\langle {\hat \phi_2 ^{\dag} (Q )\hat
\phi_2 (Q )} \right\rangle $ is approximately $ 3^{4/3} 2^{ - 1/3}
\tilde \alpha NM^{ - 1} g^{1/3} \left( {\omega _ + ^{ - 1} +
\omega _ - ^{ - 1} } \right)/32 $, which is an increasing function
of $g$. Such an analytical estimation matches the exact numerical
results in Fig. 3b with about $92 \%$ accuracy. If a higher
accuracy is desired, the ansatz (9) should be generalized. The
exact solution of a dark-soliton pair by Blow and Doran
\cite{repel} in a homogeneous background should provide insights
for making corrections.

To conclude, we have investigated the fundamental fluctuations in
dark solitons arising from their Bogoliubov quasiparticle vacuum.
These quantum fluctuations are revealed in the form of incoherent
atoms strongly concentrated near the notches. We interpret such
a quantum effect as a consequence of zero-point oscillations
of the coupled dark solitons. By constructing a particle-like
dynamical model with solitons' positions and velocities as
(finite numbers) degrees of freedom, quantitative predictions
can be made approximately. In particular we find the regime
$N^{1/6} \bar \eta ^{2/3} > 1 $ where the position uncertainty
becomes comparable with the soliton's width. Therefore
{\em blurred} dark solitons would occur if the particle number
is high or the transverse area of the condensate is sufficiently
small.

\acknowledgments We acknowledge the support from an RGC Earnmarked
Grant CUHK4237/01P and the Chinese University of Hong Kong Direct
Grant (Grant Nos: 2060148 and 2060150).

\begin{figure}
\centerline{\epsfxsize=3.3in \epsfbox{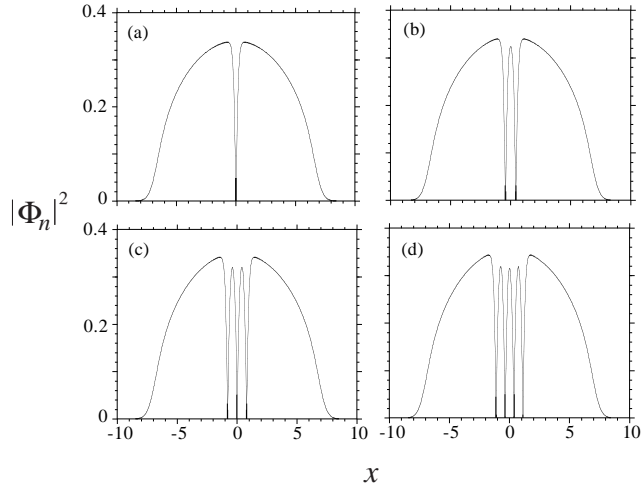}}
\vspace{5mm}
\caption{The first four stationary dark soliton states obtained
from the numerical solutions of Eq. (2) with $g=200$: (a) $n=1$,
(b) $n=2$, (c) $n=3$ and (d) $n=4$. The spatial coordinate $x$ is
in units of $\sqrt {\hbar / m \omega_T}$.}\label{fig1}
\end{figure}

\begin{figure}
\centerline{\epsfxsize=3.3in \epsfbox{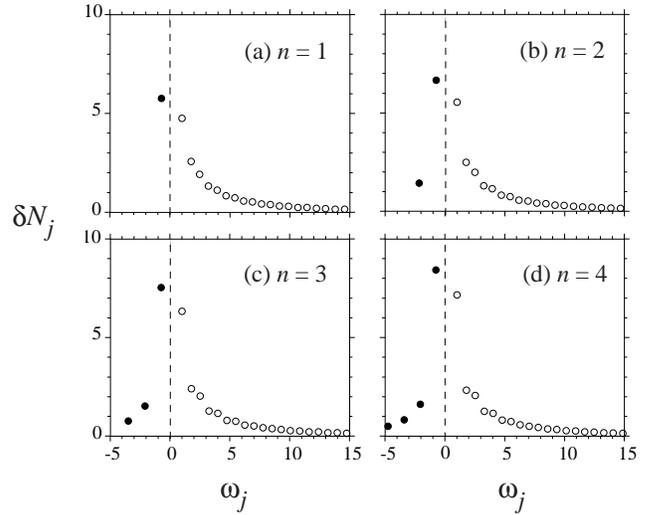}}
\vspace{5mm}

\caption{The number of incoherent atoms associated with each
collective excitation mode with the frequency $\omega_j$. The four
plots correspond to the first four dark soliton states in Fig. 1
with the same interaction strength $g=200$. The  negative
frequencies modes are anomalous modes plotted in black solid
circles. The frequencies $\omega_j$ are in units of $\omega_T$.}
\label{fig2}
\end{figure}

\begin{figure}
\centerline{\epsfxsize=3.3in \epsfbox{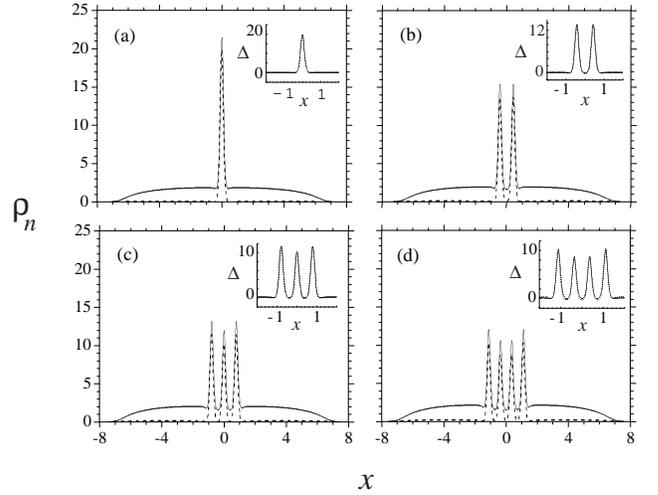}}
\vspace{5mm}
\caption{Spatial density profiles of incoherent atoms (solid
lines): (a) $n=1$, (b) $n=2$, (c) $n=3$ and (d) $n=4$ with
the same parameters used in Fig. 1. The dash
line in each figure indicates the contribution from the anomalous
modes only. The solid line in each inset shows the quantity
$\Delta = \rho_n (x) - \rho_0 (x)$ as a function of $x$.  In each
inset we have also plotted the contribution from the anomalous
modes in a dash line (that appears almost exactly the same as
$\Delta$).} \label{fig3}
\end{figure}


\begin{references}
\bibitem{eta} For a cigar shape condensate with an average
transverse area $S$ and $s$-wave scattering length $a$, the
dimensionless interaction strength $\bar \eta$ is $ 4\pi aL/S$
with $L=\left( {\hbar /m\omega_T } \right)^{1/2} $.
\bibitem{zero} Dark solitons are excited states which do not
exist at $T=0$K in strict sense. We use the word `zero-point' in
order to emphasize the analogy with the intrinsic quantum
uncertainties familiar in harmonic oscillators.
\bibitem{remark}To obtain Eq. (12), we made a
Taylor expansion of $\tilde \Phi _2$ in a power series of
$q_{\pm}$ and $ p_{\pm}$. The linear terms approximate the part of
$\hat \phi_2$ with anomalous modes, after treating $q_{\pm}$ and
$p_{\pm}$ as operators.
\end{references}
\end{document}